\documentclass[review]{elsarticle}

\usepackage{lineno,hyperref}
\modulolinenumbers[5]

\usepackage[version=4]{mhchem}
\usepackage{threeparttable}
\begin{document}
\newcommand{\ib}[1]{{\textbf#1}}
\newcommand{\Gl}{Eq.}
\newcommand{\gl}{eq.}
\newcommand{\Gls}{Eqs.}
\newcommand{\gls}{eqs.}
\newcommand{\figname}{Figure~}
\newcommand{\figsname}{Figures~}

\begin{frontmatter}
\title{Critical analysis of radical scavenging properties of atorvastatin in methanol recently estimated via density functional theory}

\author{Ioan B\^aldea}
\address{Theoretical Chemistry, Heidelberg University, Im Neuenheimer Feld 229, D-69120 Heidelberg, Germany}
\ead{ioan.baldea@pci.uni-heidelbeg.de}

\begin{abstract}
  In this communication we draw attention on serious flaws that plague recently reported antioxidant properties of
  atorvastatin (ATV) in methanol. First and foremost, we emphasize that the 
  O-H bond dissociation energies (BDE) of about 400\,kcal/mol previously reported are completely wrong.
  Further, we present results refuting the previous claim that the proton affinity (PA) of ATV is smaller
  than that of the ascorbic acid. That unfounded claim relies on
  incorrect data for PA's ascorbic acid (which we correct here) circulated in the literature.
  Further, we correct the values of the chemical reactivity indices
  (e.g., chemical hardness, electrophilicity index, electroaccepting and electrodonating
  powers), which were inadequately estimated previously via Kohn-Sham HOMO and LUMO energies.
  Finally, our updated values for \ce{O-H} bond dissociation enthalpy (BDE = 91.4\,kcal/mol) and
  electron transfer enthalpy (ETE = 105.7\,kcal/mol) tentatively suggest that
  direct H-atom transfer (HAT) and sequential proton loss electron transfer (SPLET) may coexist.
\end{abstract}


\end{frontmatter}

\section{Introduction}

Atorvastatin (ATV) \cite{Roth:02} is one of the best selling drug
belonging to the class of statins \cite{Istvan:01,Istvan:02}
widely administered over more than 25 years in a variety of
settings to lower the ``bad'' cholesterol and fats ---e.g., low-density lipoprotein (LDL), triglycerides ---,
decrease the risk of heart disease, prevent strokes and heart attacks, etc \cite{Vuppalanchi:06,Ingold:14,Alenghat:19,Galano:19}.

Quantum chemistry can provide important insight into the antioxidant activity of ATV, which still remained elusive.
For this reason, the first efforts in this direction undertaken only recently \cite{Duque:22}
are certainly welcome. Unfortunately, the results presented in ref.~\cite{Duque:22}
are plagued by serious flaws, and drawing attention to this fact is an important aim of the
present communication. As brief justification of this assertion suffice it to mention what is
undoubtedly the most eye-catching error of ref.~\cite{Duque:22}. For ATV's \ce{O-H} and \ce{N-H} bonds, 
values of about 400\,kcal/mol ($>17$\,eV) are reported (cf.~Table 2 of ref.~\cite{Duque:22}).
Roughly, these are four times larger
than any BDE value characterizing the well documented \ce{O-H} and \ce{N-H} groups \cite{Ingold:14}, 
which represent mandatory ingredients of any radical scavenger just because of their well known
low BDE's. The aforementioned values even by far exceed the strongest chemical bonds known today,
amounting to (generously speaking) $\sim 10$\,eV. 
\section{Computational details}
The quantum chemical calculations based on the density functional theory (DFT)
done in conjunction with this study employed
the GAUSSIAN 16 \cite{g16} suite of programs on the bwHPC platform \cite{bwHPC2}.
We used the three parameter B3LYP
hybrid DFT/HF exchange correlation functional \cite{Parr:88,Becke:88,Becke:93a,Frisch:94}
along with the Pople 6-31+G(d,p) basis sets \cite{Petersson:88,Petersson:91}.
Solvent (specifically, methanol) effects were accounted for
within the polarized continuum model (PCM) \cite{Tomasi:05} using
the integral equation formalism (IEF) \cite{Cances:97}.

For open shell radicals 
($\ce{ATV^{\bullet +}}$ and
\ce{ATVwoH$^{\bullet}$}, cf.~Tables~\ref{table:energies}, \ref{table:ATV+},
and \ref{table:ATV1H}) we performed unrestricted (UB3LYP) calculations.
Therefore, it is worth mentioning that, similar to other 
similar cases \cite{Baldea:2019e,Baldea:2019g,Baldea:2022d,Baldea:2022e},
we can safely rule out spin contamination artifacts (cf.~Table~\ref{table:energies}).

Because literature
studies on antioxidant activity often use optimized geometries and vibrational corrections due to
zero point motion in the gas phase with smaller basis sets, and only compute electronic energies
with solvent and larger basis sets, we emphasize that all our optimized
geometries (presented in Tables~\ref{table:ATV} to \ref{table:vitC}) and
vibrational corrections due to zero point motion
(Table~\ref{table:energies})
were also computed at the B3LYP/6-31+G(d,p)/IEFPCM level of theory.
GABEDIT \cite{gabedit} was used to generate \figsname\ref{fig:ATV} and \ref{fig:vitC}.
All thermodynamic properties were calculated at $T = 298.15$\,K.
\section{Results and discussion}
Antioxidants ({X}H) are molecules that inhibit oxidation processes by transferring a hydrogen atom
to free radicals (R). The three main antioxidative mechanisms (HAT, SET-PT, and SPLET)
along with the pertaining reaction enthalpies (BDE, IP and PDE, PA and ETE, respectively) are depicted below.

Direct hydrogen atom transfer (HAT):
\begin{equation}
  \label{eq-hat}
  \begin{array}{cc}
    \ce{{X}H + R^{\bullet}} \to \ce{{X}^{\bullet}} + \ce{RH} &
        \mbox{\small BDE} = H\left(\ce{{X}^{\bullet}}\right) + H\left(\ce{H^{\bullet}}\right) - H\left(\ce{{X}H}\right) 
  \end{array}
\end{equation}
Stepwise electron transfer proton transfer (SET-PT):
\begin{subequations}
  \label{eq-setpt}
\begin{equation}
  \label{eq-setpt-ip}
  \begin{array}{cc}
    \ce{{X}H} \to \ce{{X}H^{\bullet +}} + \ce{e-} &
    \mbox{\small IP} = H\left(\ce{{X}H^{\bullet +}}\right) + H\left(\ce{e-}\right) - H\left(\ce{{X}H}\right) 
     \end{array}
\end{equation}
\begin{equation}
    \label{eq-setpt-pde}
  \begin{array}{cc}
    \ce{{X}H^{\bullet +}} \to  \ce{{X}^{\bullet}} + \ce{H+} &
      \mbox{\small PDE} = H\left( \ce{{X}^{\bullet}}\right) + H\left(\ce{H+}\right) -  H\left(\ce{{X}H^{\bullet +}}\right)      
  \end{array}
\end{equation}
\end{subequations}
Sequential proton loss electron transfer (SPLET):
\begin{subequations}
    \label{eq-splet}
\begin{equation}
    \label{eq-splet-pa}
  \begin{array}{cc}
    \ce{{X}H} \to \ce{{X}-} + \ce{H+} &
     \mbox{\small PA} = H\left(\ce{{X}-}\right) + H\left(\ce{H+}\right) - H\left(\ce{{X}H}\right) 
     \end{array}
\end{equation}
\begin{equation}
      \label{eq-splet-ete}
  \begin{array}{cc}
    \ce{{X}-} \to  \ce{{X}^{\bullet}} + \ce{e-} &
      \mbox{\small ETE} = H\left( \ce{{X}^{\bullet}}\right) + H\left(\ce{e-}\right) -  H\left(\ce{{X}-}\right)      
  \end{array}
\end{equation}
\end{subequations}

The thermodynamic parameters listed above (BDE, IP and PDE, PA and ETE) provide 
useful information needed to assess the radical scavenging activity of ATV. They are enthalpies of reaction
and can be obtained from standard $\Delta$-DFT quantum chemical calculations \cite{Gunnarson:89,Baldea:2012i,Baldea:2014c}).

\begin{figure} 
  \centerline{\includegraphics[width=0.45\textwidth,angle=0]{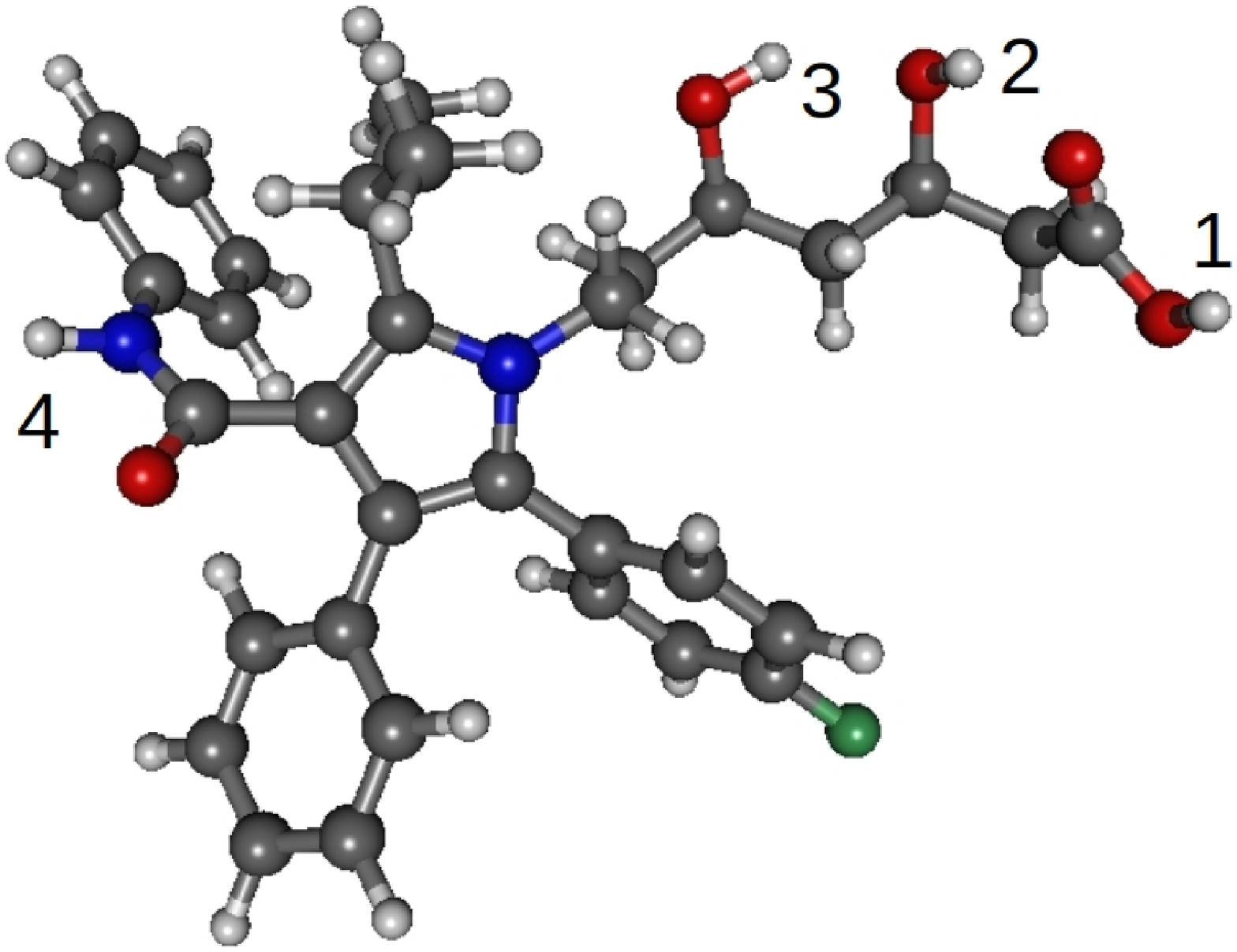} 
    \includegraphics[width=0.45\textwidth,angle=0]{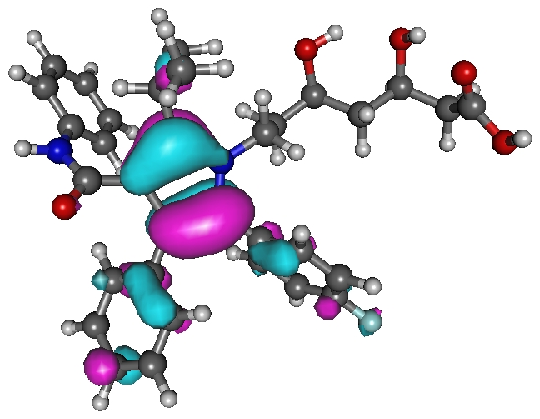}
    \includegraphics[width=0.45\textwidth,angle=0]{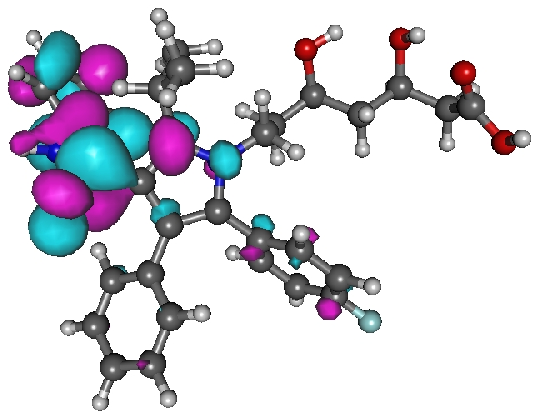}
    }
  \caption{Optimized geometry with labels of the O-H groups (left), HOMO (middle), and LUMO (right)
    spatial distributions of the atorvastatin molecule (ATV).}
  \label{fig:ATV}
\end{figure}

In this communication, we will restrict ourselves to
the O-H bond of the carboxylic acid group (1-OH position in \figname\ref{fig:ATV}),
which, out of the three O-H bonds of ATV, has according to ref.~\cite{Duque:22}
the lowest PA and ETE. 
In the context of ref.~\cite{Duque:22}, the lowest ATV's PA value  
is not only relevant for the SPLET mechanism (claimed to be the most efficient pathway in ATV \cite{Duque:22})
but also because of the claim \cite{Duque:22} that it is lower than that of ascorbic acid (vitamin C).
To refute this claim, we also re-estimated PA for ascorbic acid
and emphasize the difference from an inadequate value circulated in the literature.
Unlike that value shown in ref.~\cite{Duque:22},
our PA for ascorbic acid was computed at the same
B3LYP/6-31+G(d,p)/IEFPCM level of theory also employed for ATV. 

The presently calculated enthalpies of the ATV species entering the right-hand side
of \gls~(\ref{eq-hat}), (\ref{eq-setpt}), and (\ref{eq-splet})
are included in Table~\ref{table:energies}.
To estimate the various reaction energies,
the gas phase enthalpies of the hydrogen atom, proton, and electron  
as well as their solvation enthalpies are also needed. They are listed in Table~\ref{table:values}. 
\begin{table} 
  \scriptsize 
  \begin{center}
    \begin{tabular*}{0.85\textwidth}{@{\extracolsep{\fill}}lcccccc}
      \hline
      Species &    $E_0$    &   $ZPE$   &  $TCH$ & $\left\langle S^2\right\rangle_{b}$ & $\left\langle S^2\right\rangle_{a}$ \\
      \hline
      ATV     & -1864.24567013 &  0.623552 & 0.663158 &        &        \\
 $\ce{ATV^{\bullet +}}$ & -1864.04563713 &  0.624017 & 0.663523 & 0.7633 & 0.7501 \\
      $\ce{ATVwoH^{\bullet}}$   & -1863.59072281 &  0.610822 & 0.649807 & 0.7633 & 0.7501 \\
      \ce{ATVwoH-}  & -1863.78861771 &  0.610362 & 0.649465 &        &        \\
      \hline
    \end{tabular*}
    \caption{Electronic energy $E_0$, zero point vibrational energy $ZPE$, and thermal corrections to enthalpy $TCH$
      for neutral ATV and related species in methanol at the B3LYP/6-31+G(d,p) level of theory. For open shell systems, calculations were done
        within the unrestricted (UB3LYP) framework, and the values of the total spin before (label $b$) and after (label $a$)
        annihilation of the first spin contaminant are indicated.
        $\ce{ATV^{\bullet +}}$ = ATV cation; $\ce{ATVwoH^{\bullet}}$ = ATV radical after H-atom removal from 1-OH position
        (cf.~\figname\ref{fig:ATV}).
    }
  \label{table:energies}
\end{center}
\end{table}
\begin{table} 
  \scriptsize 
  \begin{center}
 \begin{threeparttable}
    \begin{tabular*}{0.5\textwidth}{@{\extracolsep{\fill}}lcccccc}
      \hline
      Species  &       $H_0$      &  $\Delta H_{sol}^{methanol}$ \\ 
      \hline
      Electron & +0.001194\tnote{$^a$}\hspace*{-1.5ex}      & -0.030204\tnote{$^c$}\hspace*{-1.5ex}  \\
      Proton   & +0.002339\tnote{$^b$}\hspace*{-1.5ex}      & -0.405508\tnote{$^c$}\hspace*{-1.5ex}  \\
      H-atom   & -0.500273                                  & +0.001904\tnote{$^d$}\hspace*{-1.5ex}  \\
      \hline
    \end{tabular*}
    \begin{tablenotes} 
    \item[$^a$] From Ref.~\citenum{Fifen:13}
    \item[$^b$] From Ref.~\citenum{Fifen:14}
    \item[$^c$] Form Ref.~\citenum{Markovic:16}
    \item[$^d$] Form Ref.~\citenum{Rimarcik:10} 
    \end{tablenotes}
  \end{threeparttable}
    \caption{Gas phase enthalpies $H_0$ and solvation enthalpies $\Delta H_{sol}$ 
      in hartree needed in the present calculations. Except for the gas phase enthalpy 
      of the hydrogen atom computed by us using B3LYP/6-31+G(d,p),
      all other entries are literature data \cite{Rimarcik:10,Fifen:13,Fifen:14,Markovic:16}.} 
  \label{table:values}
\end{center}
\end{table}

\begin{figure} 
  \centerline{\includegraphics[width=0.3\textwidth,angle=0]{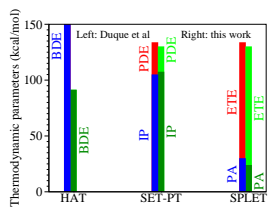}}   
  \caption{Thermodynamic parameters quantifying the radical scavenging activity of ATV in methanol as estimated in the
  present study along with those reported in ref.~\cite{Duque:22}. The underlying numerical values were taken from Table~\ref{table:bde}.}
  \label{fig:bde}
\end{figure}

Insertion of the values of Tables~\ref{table:values} and \ref{table:energies} into
\gls~(\ref{eq-hat}), (\ref{eq-setpt}), and (\ref{eq-splet})
yields the desired ATV's thermodynamic parameters.
They are given in Table~\ref{table:bde} and visualized in \figname\ref{fig:bde}
along with the previous values (written in italics) of ref.~\cite{Duque:22}.
As visible there, in contrast to the huge value (403.8\,kcal/mol),
our value for ATV's O-H BDE (91.4\,kcal/mol) has nothing extravagant;
it lies in the range typical for hydroxy groups \cite{Ingold:14}.

Although neither the finding on incorrect BDE value previously reported \cite{Duque:22}
nor our updated BDE value should be too surprising for scholars familiar with O-H groups,
the aforementioned is also important because it challenges the claim of ref.~\cite{Duque:22}
that SPLET is the dominant mechanism responsible for the radical scavenging activity of ATV.
This claim on the SPLET prevalence is hardly substantiated by the updated value BDE$=91.4$\,kcal/mol,
``comfortably'' lower than ETE=$106.2$\,kcal/mol (second step of SPLET, cf.~\gl~(\ref{eq-splet-ete}))
and also lower than IP=$107.0$\,kcal/mol (first step of SET-PT, cf.~\gl~(\ref{eq-setpt-ip})).
In view of our results (Table~\ref{table:bde} and \figname\ref{fig:bde}), we suggest that
HAT rather than SPLET is the ATV's dominant antioxidant mechanism in methanol.

Our proton affinity (PA=$23.8$\,kcal/mol, cf.~Table~\ref{table:bde})
is even lower than the previous estimate (PA=$30.1$\,kcal/mol \cite{Duque:22}).
Nevertheless, as elaborated below, we
refute the claim \cite{Duque:22} that this value is smaller than for ascorbic acid.

\begin{figure} 
  \centerline{\includegraphics[width=0.3\textwidth,angle=0]{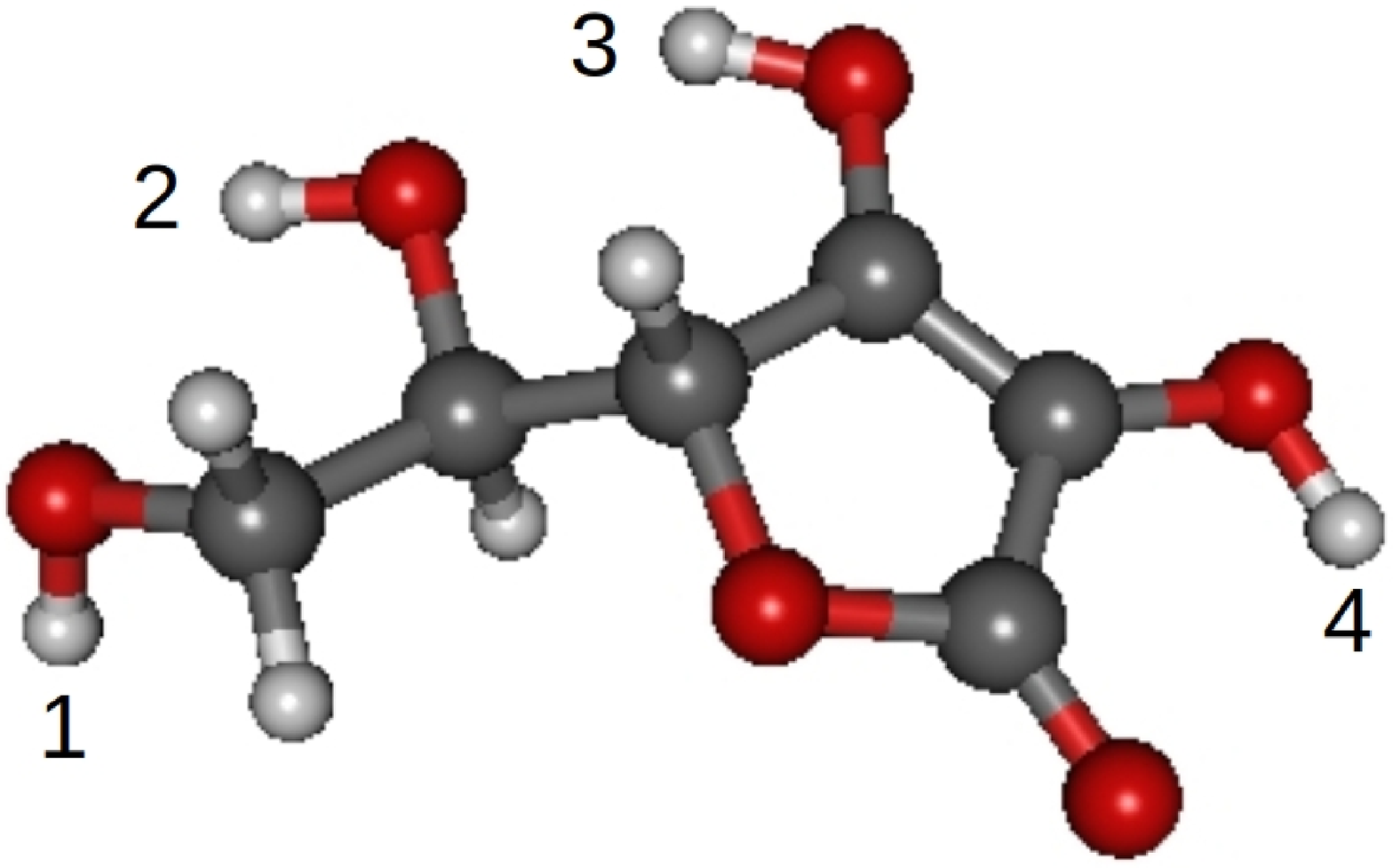}     
    \includegraphics[width=0.3\textwidth,angle=0]{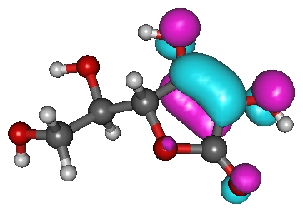}
    \includegraphics[width=0.3\textwidth,angle=0]{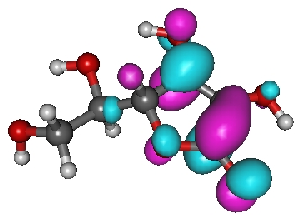}
  }
  \caption{Optimized geometry with labels of the O-H groups (left), HOMO (middle), and LUMO (right)
    spatial distributions of ascorbic acid.}
  \label{fig:vitC}
\end{figure}

To support that claim, ref.~\cite{Duque:22} quoted a value PA=$34.2$\,kcal/mol for ascorbic acid
taken from literature but did not mention to which O-H group this value belongs. This is a significant issue
because there are four O-H groups in ascorbic acid (cf.~\figname\ref{fig:vitC}).
To elucidate this point, we also performed B3LYP/6-31+G(d,p)/IEFPCM calculations for all the four
O-H groups of ascorbic acid in methanol.
Our results (to be reported in detail elswhere) settles the confusion.
Up to minor difference understandable in view of the different levels of theory,
the value of PA=$34.2$\,kcal/mol for ascorbic acid presented in ref.~\cite{Duque:22} corresponds to the 4-OH position (cf.~\figname\ref{fig:vitC}),
for which we estimated PA=$35.0$\,kcal/mol.
However, this is not the lowest PA of ascorbic acid. The lowest PA of ascorbic acid determined by us
(20.5\,kcal/mol) corresponds to the 3-OH position.

A detailed comparison between our data for ascorbic acid and those
from previous literature including ref.~\cite{Duque:22} deserve separate consnideration. 
What matters from the present discussion is that the lowest PA value (20.5\,kcal/mol) of ascorbic acid
is lower than the ATV's lowest value (PA=$23.8$\,kcal/mol) presently considered. 

\begin{table} 
  \scriptsize 
  \begin{center}
    \begin{tabular*}{0.85\textwidth}{@{\extracolsep{\fill}}rr|rr|rr}
      \hline
                        &   HAT          &   SET-PT     &               &    SPLET      &            \\    
      \hline
                        & BDE            &         IP   & PDE            &        PA    &       ETE   \\  
      \hline                                                                           
  Present work                 &         91.4   &       107.0  &        22.4    &        23.8  &       105.7  \\
  Ref.~\cite{Duque:22}  &  \emph{403.8}  & \emph{105.1} &  \emph{28.8}   &  \emph{30.1} & \emph{103.8} \\
     \hline                                                                                                                 
    \end{tabular*}
    \caption{\ce{O-H} bond dissociation enthalpies (BDE), ionization enthalpies (IP), proton dissociation enthalpies (PDE),
      proton affinity (PA), and electron transfer enthalpies (ETE) computed for ATV and ascorbic acid in methanol at the
      B3LYP/6-31+G(d,p) level of theory. Values previously reported in ref.~\cite{Duque:22} are shown in italics.
      All quantities in kcal/mol.
    }
  \label{table:bde}
\end{center}
\end{table}

Ref.~\cite{Duque:22} also reported global chemical reactivity indices for ATV in methanol:
chemical hardness $\eta \equiv E_g/2$, chemical softness $S \equiv 1/E_g$,
electronegativity $\chi \equiv (\mbox{IP} + \mbox{EA})/2$, electrophilicity index
$\omega \equiv \chi^2/(2 \eta)$
as well as electroaccepting 
$\omega^{+} \equiv  (\mbox{IP} + 3\, \mbox{EA})^2/(16 E_g) $
and electrodonating 
$\omega^{-} \equiv (3 \,\mbox{IP} + \mbox{EA})^2/(16 E_g) $ powers.
Here, $E_g \equiv  \mbox{IP} - \mbox{EA}$ is the fundamental (or transport) ``HOMO-LUMO''
gap \cite{Parr:89,Burke:12,Baldea:2014c}. 
What makes their estimates of Ref.~\cite{Duque:22} highly questionable 
is the determination of the ionization energy (IP) and electroaffinity (EA)
from the Kohn-Sham (KS) HOMO and LUMO energies obtained from DFT calculations in solvent
($\mbox{IP} \approx I \equiv - E_{HOMO}^{KS}$, $\mbox{EA} \approx A \equiv - E_{LUMO}^{KS}$).
It is well known that the KS gap $I - A$ differs from $E_g$ even if
computed with the exact exchange-correlation functional \cite{Perdew:82,Burke:12},
and the presence of a solvent adds a further difficulty \cite{Baldea:2013c}.

To estimate the global chemical reactivity indices, we computed the electroaffiity
$ \mbox{EA} = H\left(\ce{{X}H}\right) + H\left(\ce{e-}\right) - H\left(\ce{{X}H^{\bullet -}}\right)$
from the enthalpy of the anion
($\ce{{X}H^{\bullet -}} \equiv \ce{ATV^{\bullet -}}$),
i.e., as counterpart of
of IP of \gl~(\ref{eq-setpt-ip}). As expected, our values collected in Table~\ref{table:eta}
substantially differ from those previously reported \cite{Duque:22}.
\begin{table} 
  \scriptsize 
  \begin{center}
    \begin{tabular*}{1.0\textwidth}{@{\extracolsep{\fill}}rccccccccc}
      \hline
                      &       IP     &        EA   & $E_g$       & $\chi \equiv - \mu$ &     $\eta$   &      $S$    &  $\omega$   & $\omega^{+}$ & $\omega^{-}$ \\  
      \hline                                                                           
 Present work         &  4.64        &  0.72       &   3.92      & 1.96        &  -2.68       & 0.26        & 1.83        &      0.74   & 3.42        \\
 Ref.~\cite{Duque:22} &  \emph{5.75} & \emph{1.04} & \emph{4.71} & \emph{2.35} & \emph{-3.40} & \emph{0.21} & \emph{2.45} & \emph{1.05} & \emph{4.44} \\
      \hline                                                                           
    \end{tabular*}
    \caption{Presently computed global chemical reactivity indices (in eV) substantially differ from those estimated in ref.~\cite{Duque:22} shown in italics.
    }
  \label{table:eta}
\end{center}
\end{table}
\section*{Conclusion}
In closing, clearly contradicting the main conclusions of ref.~\cite{Duque:22},
the results presented above demonstrated that the O-H bonds of atorvastatin (ATV) are by no means unusually strong,
and that the lowest proton affinity of ascorbic acid is lower than the lowest proton affinity of ATV.
In addition, we showed that the global (adiabatic) chemical reactivity indices based on IP and EA values 
reliably determined via enthalpies of reactions substantially differ from those based
on the Kohn-Sham HOMO and LUMO energies.

Last but not least, the presently corrected value of the \ce{O-H} bond dissociation energy, smaller than the second
step of the SPLET (BDE = 91.4\,kcal/mol versus ETE = 105.7\,kcal/mol), may tentatively suggest
a bottleneck effect played by the second step of the SPLET pathway and herewith 
a nontrivial interplay with direct hydrogen atom transfer (HAT) 
in determining the antioxidant activity of ATV in methanol.
\section*{Acknowledgments}

The author thanks Ederley V\'elez Ortiz for the kindness in providing him valuable details related to
her recent work \cite{Duque:22}.
Financial support from the German Research Foundation
(DFG Grant No. BA 1799/3-2) in the initial stage of this work and computational support by the
state of Baden-W\"urttemberg through bwHPC and the German Research Foundation through
Grant No.~INST 40/575-1 FUGG 
(bwUniCluster 2.0, bwForCluster/MLS\&WISO 2.0, and JUSTUS 2.0 cluster) are gratefully acknowledged.

\newpage
\centerline{\textbf{Appendix}}
\renewcommand{\theequation}{S\arabic{equation}}
\setcounter{equation}{0}
\renewcommand{\thefigure}{S\arabic{figure}}
\setcounter{figure}{0}
\renewcommand{\thetable}{S\arabic{table}}
\setcounter{table}{0}
\renewcommand{\thesection}{S\arabic{section}}
\setcounter{section}{0}
\renewcommand{\thepage}{S\arabic{page}}
\setcounter{page}{0}
\renewcommand{\thefootnote}{\alph{footnote}}

\begin{table} 
\small 
  \begin{center}
    \caption{Coordinates of ATV optimized in methanol via RB3LYP/6-31+G(d,p)/IEFPCM. All values in angstrom.}
    \begin{tabular*}{0.95\textwidth}{@{\extracolsep{\fill}}rrrr|rrrr}
Atom      &                    X        &         Y         &          Z       &   Atom      &                    X        &         Y         &          Z       \\
\hline
 F        &              -3.10161200    &     4.92704700    &    -1.32781200   &    C        &              -0.32790500    &     2.57709400    &    -1.35193100   \\[-0.3ex]    
 O        &              -6.57690300    &    -1.23310100    &     1.55261700   &    H        &               0.44269000    &     2.22333700    &    -2.02930000   \\[-0.3ex]    
 O        &               4.41731700    &     0.32594500    &     2.25091400   &    C        &              -1.22961800    &     3.54944900    &    -1.79029700   \\[-0.3ex]    
 N        &               0.25030100    &    -0.25509300    &     0.75806400   &    H        &              -1.17733000    &     3.96041800    &    -2.79273000   \\[-0.3ex]    
 C        &              -6.82160200    &    -0.44128000    &     0.65138800   &    C        &              -2.32369000    &     3.48862800    &     0.38698700   \\[-0.3ex]    
 C        &               3.86938900    &    -0.35311100    &     1.36852800   &    H        &              -3.09917400    &     3.86083200    &     1.04770900   \\[-0.3ex]    
 C        &               1.39000000    &    -0.96319700    &     1.09384600   &    C        &              -1.40634700    &     2.52375600    &     0.80931800   \\[-0.3ex]    
 C        &               0.59359200    &     1.04694000    &     0.40288400   &    H        &              -1.46479800    &     2.15264100    &     1.82800500   \\[-0.3ex]    
 C        &              -2.21114900    &     3.97957500    &    -0.90732300   &    C        &              -1.10282800    &    -0.80924800    &     0.60291400   \\[-0.3ex]    
 C        &               4.33941000    &    -2.19429600    &    -0.34298900   &    H        &              -1.81198800    &    -0.06894900    &     0.97457100   \\[-0.3ex]    
 C        &               2.46907000    &    -0.09340200    &     0.94903000   &    H        &              -1.19074400    &    -1.68262200    &     1.24364900   \\[-0.3ex]    
 C        &               1.96955300    &     1.18500800    &     0.52827000   &    C        &              -3.98389200    &    -0.99399800    &    -0.65294500   \\[-0.3ex]    
 C        &              -0.40019100    &     2.04821300    &    -0.05111300   &    H        &              -3.88834100    &    -0.02560700    &    -1.15958400   \\[-0.3ex]    
 C        &               2.76038900    &     2.41096900    &     0.27950500   &    H        &              -3.97630900    &    -0.80376900    &     0.42705600   \\[-0.3ex]    
 C        &               0.77009300    &    -2.42549900    &     3.08542600   &    C        &              -6.53122200    &    -0.68786500    &    -0.81020800   \\[-0.3ex]    
 H        &               1.19347100    &    -1.66684000    &     3.75129200   &    H        &              -7.43468400    &    -1.14786400    &    -1.23125000   \\[-0.3ex]    
 H        &              -0.31348200    &    -2.27576900    &     3.05035100   &    H        &              -6.39312900    &     0.26697000    &    -1.32380100   \\[-0.3ex]    
 H        &               0.95182800    &    -3.40966800    &     3.53092400   &    O        &              -2.75986800    &    -3.10006300    &    -0.30287200   \\[-0.3ex]    
 C        &               4.88844400    &    -3.48661300    &    -0.34505400   &    H        &              -3.68170000    &    -3.41604300    &    -0.27105500   \\[-0.3ex]    
 H        &               5.43269900    &    -3.83815300    &     0.52720200   &    O        &              -5.53024400    &    -2.90105100    &    -0.43216300   \\[-0.3ex]    
 C        &               4.73504000    &    -4.31718800    &    -1.45643900   &    H        &              -5.81481400    &    -2.72152400    &     0.48344000   \\[-0.3ex]    
 H        &               5.16739000    &    -5.31325600    &    -1.44104900   &    O        &              -7.42117700    &     0.73769500    &     0.87605400   \\[-0.3ex]    
 C        &               4.02318300    &    -3.87489700    &    -2.57572800   &    H        &              -7.62466500    &     0.80861600    &     1.82669300   \\[-0.3ex]    
 H        &               3.89861600    &    -4.52186900    &    -3.43850800   &    N        &               4.57984700    &    -1.38251800    &     0.79596400   \\[-0.3ex]    
 C        &               3.48057200    &    -2.58617600    &    -2.57438000   &    H        &               5.45749000    &    -1.55457100    &     1.27532400   \\[-0.3ex]    
 H        &               2.93677500    &    -2.22360200    &    -3.44185400   &    C        &               1.42256300    &    -2.35903400    &     1.68547100   \\[-0.3ex]    
 C        &               3.64189500    &    -1.74201900    &    -1.47322700   &    H        &               2.48737100    &    -2.55783400    &     1.84145800   \\[-0.3ex]    
 H        &               3.22789500    &    -0.74191300    &    -1.49676100   &    C        &               0.91213100    &    -3.48775000    &     0.76412600   \\[-0.3ex]    
 C        &               2.37032200    &     3.64855500    &     0.82485900   &    H        &              -0.17116100    &    -3.46314100    &     0.62282600   \\[-0.3ex]    
 H        &               1.48228200    &     3.69984500    &     1.44686600   &    H        &               1.38776000    &    -3.43813800    &    -0.21976100   \\[-0.3ex]    
 C        &               3.11586300    &     4.80643400    &     0.58994200   &    H        &               1.16264100    &    -4.45635300    &     1.21043400   \\[-0.3ex]    
 H        &               2.79422300    &     5.74881400    &     1.02466500   &    C        &              -1.42112600    &    -1.18590100    &    -0.85320900   \\[-0.3ex]    
 C        &               4.27557900    &     4.75359100    &    -0.19014500   &    H        &              -0.64463300    &    -1.86432300    &    -1.22262900   \\[-0.3ex]    
 H        &               4.85736400    &     5.65285600    &    -0.37037800   &    H        &              -1.38982800    &    -0.29025500    &    -1.48271800   \\[-0.3ex]    
 C        &               4.68062300    &     3.52977600    &    -0.73262300   &    C        &              -2.78024500    &    -1.86891300    &    -1.04375000   \\[-0.3ex]    
 H        &               5.57865800    &     3.47361900    &    -1.34153900   &    H        &              -2.87280900    &    -2.10211400    &    -2.11710700   \\[-0.3ex]    
 C        &               3.93034100    &     2.37432200    &    -0.50176600   &    C        &              -5.32972000    &    -1.61629800    &    -1.05953500   \\[-0.3ex]    
 H        &               4.25330800    &     1.43572900    &    -0.94222800   &    H        &              -5.30349800    &    -1.82877300    &    -2.13351500   \\[-0.3ex]    
    \hline                                                                                                                                                          
    \end{tabular*}                                                                                                                                                  
  \label{table:ATV}                                                                                                                                                 
\end{center}                                                                                                                                                        
\end{table}                                                                                                                                                         
\begin{table} 
\small 
  \begin{center}                                                                                                                                                    
    \caption{Coordinates of $\ce{ATV^{\bullet +}}$ radical cation optimized in methanol via UB3LYP/6-31+G(d,p)/IEFPCM. All values in angstrom.}                     
    \begin{tabular*}{0.95\textwidth}{@{\extracolsep{\fill}}rrrr|rrrr}                                                                                               
Atom      &         X        &      Y        &          Z       & Atom    &         X        &      Y        &         Z       \\                                   
\hline
 F        &       -2.544038  &     5.259388  &    -1.397056     &    C    &       0.131406   &    2.828285   &   -1.232676     \\[-0.3ex]    
 O        &       -6.630600  &    -1.027423  &     1.570355     &    H    &       0.997283   &    2.521115   &   -1.807518     \\[-0.3ex]    
 O        &        4.391284  &    -0.057308  &     2.298101     &    C    &      -0.663418   &    3.867514   &   -1.703098     \\[-0.3ex]    
 N        &        0.205551  &    -0.184885  &     0.782218     &    H    &      -0.431193   &    4.383314   &   -2.627703     \\[-0.3ex]    
 C        &       -6.894429  &    -0.249615  &     0.662324     &    C    &      -2.123976   &    3.615475   &    0.243017     \\[-0.3ex]    
 C        &        3.846333  &    -0.635431  &     1.352445     &    H    &      -2.983344   &    3.959546   &    0.807057     \\[-0.3ex]    
 C        &        1.295381  &    -0.987486  &     1.106190     &    C    &      -1.339967   &    2.558674   &    0.690408     \\[-0.3ex]    
 C        &        0.651578  &     1.056695  &     0.417589     &    H    &      -1.581913   &    2.092796   &    1.638661     \\[-0.3ex]    
 C        &       -1.773677  &     4.239793  &    -0.951944     &    C    &      -1.185210   &   -0.667053   &    0.607185     \\[-0.3ex]    
 C        &        4.091951  &    -2.438099  &    -0.424394     &    H    &      -1.856546   &    0.093024   &    0.997898     \\[-0.3ex]    
 C        &        2.469539  &    -0.223356  &     0.919347     &    H    &      -1.307618   &   -1.546749   &    1.230729     \\[-0.3ex]    
 C        &        2.099035  &     1.065834  &     0.523379     &    C    &      -4.061815   &   -0.801008   &   -0.657196     \\[-0.3ex]    
 C        &       -0.203557  &     2.145058  &    -0.040532     &    H    &      -3.971836   &    0.161481   &   -1.175716     \\[-0.3ex]    
 C        &        2.964472  &     2.223721  &     0.306531     &    H    &      -4.048939   &   -0.599689   &    0.420817     \\[-0.3ex]    
 C        &        0.558944  &    -2.354523  &     3.101888     &    C    &      -6.611839   &   -0.508671   &   -0.798711     \\[-0.3ex]    
 H        &        1.029930  &    -1.616543  &     3.757256     &    H    &      -7.515687   &   -0.976545   &   -1.210030     \\[-0.3ex]    
 H        &       -0.511970  &    -2.140876  &     3.057594     &    H    &      -6.481826   &    0.441881   &   -1.322326     \\[-0.3ex]    
 H        &        0.683890  &    -3.342842  &     3.553977     &    O    &      -2.834991   &   -2.905620   &   -0.303086     \\[-0.3ex]    
 C        &        4.413724  &    -3.803653  &    -0.417301     &    H    &      -3.757839   &   -3.218900   &   -0.255863     \\[-0.3ex]    
 H        &        4.901516  &    -4.235927  &     0.451473     &    O    &      -5.595224   &   -2.712582   &   -0.404943     \\[-0.3ex]    
 C        &        4.102379  &    -4.601597  &    -1.518764     &    H    &      -5.877165   &   -2.525507   &    0.510141     \\[-0.3ex]    
 H        &        4.357338  &    -5.656723  &    -1.502024     &    O    &      -7.510438   &    0.922246   &    0.877661     \\[-0.3ex]    
 C        &        3.457218  &    -4.050133  &    -2.630688     &    H    &      -7.710329   &    1.000579   &    1.828503     \\[-0.3ex]    
 H        &        3.209781  &    -4.672771  &    -3.484604     &    N    &       4.471147   &   -1.664261   &    0.709696     \\[-0.3ex]    
 C        &        3.143422  &    -2.688601  &    -2.637800     &    H    &       5.337137   &   -1.937498   &    1.164242     \\[-0.3ex]    
 H        &        2.658512  &    -2.244633  &    -3.501789     &    C    &       1.225119   &   -2.367246   &    1.702936     \\[-0.3ex]    
 C        &        3.469010  &    -1.876913  &    -1.547734     &    H    &       2.270470   &   -2.643117   &    1.865697     \\[-0.3ex]    
 H        &        3.249646  &    -0.817250  &    -1.586653     &    C    &       0.632122   &   -3.459429   &    0.780953     \\[-0.3ex]    
 C        &        2.591308  &     3.507522  &     0.766566     &    H    &      -0.444916   &   -3.360914   &    0.633897     \\[-0.3ex]    
 H        &        1.654633  &     3.641974  &     1.295001     &    H    &       1.120036   &   -3.458528   &   -0.197264     \\[-0.3ex]    
 C        &        3.440894  &     4.596965  &     0.588768     &    H    &       0.816821   &   -4.431401   &    1.247972     \\[-0.3ex]    
 H        &        3.147227  &     5.572466  &     0.962992     &    C    &      -1.498004   &   -1.000310   &   -0.858986     \\[-0.3ex]    
 C        &        4.667620  &     4.434076  &    -0.063368     &    H    &      -0.727161   &   -1.675231   &   -1.245171     \\[-0.3ex]    
 H        &        5.324639  &     5.285997  &    -0.208009     &    H    &      -1.467455   &   -0.091799   &   -1.468952     \\[-0.3ex]    
 C        &        5.049425  &     3.167865  &    -0.526203     &    C    &      -2.860874   &   -1.679047   &   -1.047550     \\[-0.3ex]    
 H        &        5.997832  &     3.037244  &    -1.037367     &    H    &      -2.950733   &   -1.911929   &   -2.120551     \\[-0.3ex]    
 C        &        4.213747  &     2.072520  &    -0.334872     &    C    &      -5.408080   &   -1.433700   &   -1.047524     \\[-0.3ex]    
 H        &        4.518773  &     1.101287  &    -0.708739     &    H    &      -5.388469   &   -1.657067   &   -2.119338     \\[-0.3ex]    
    \hline 
    \end{tabular*}
  \label{table:ATV+}
\end{center}
\end{table}
\begin{table} 
\small 
  \begin{center}
    \caption{Coordinates of $\ce{ATVwoH^{\bullet}}$ radical optimized in methanol via UB3LYP/6-31+G(d,p)/IEFPCM. All values in angstrom.}
    \begin{tabular*}{0.95\textwidth}{@{\extracolsep{\fill}}rrrr|rrrr}
Atom      &        X        &        Y        &         Z       &   Atom    &         X        &      Y       &      Z     \\
\hline
   F      &     -2.394990   &      5.382882   &     -1.404039   &    C      &       0.199618   &   2.866108   &  -1.236101 \\[-0.3ex]    
   O      &     -6.730500   &     -1.249652   &      1.509079   &    H      &       1.060689   &   2.535662   &  -1.805272 \\[-0.3ex]    
   O      &      4.324321   &     -0.170699   &      2.325261   &    C      &      -0.557101   &   3.934121   &  -1.704682 \\[-0.3ex]    
   N      &      0.157056   &     -0.163816   &      0.756034   &    H      &      -0.300124   &   4.449936   &  -2.622714 \\[-0.3ex]    
   C      &     -6.909301   &     -0.191253   &      0.810395   &    C      &      -2.043573   &   3.712675   &   0.225733 \\[-0.3ex]    
   C      &      3.774452   &     -0.729095   &      1.370555   &    H      &      -2.897571   &   4.078419   &   0.784158 \\[-0.3ex]    
   C      &      1.216871   &     -1.001655   &      1.088124   &    C      &      -1.298080   &   2.627545   &   0.670910 \\[-0.3ex]    
   C      &      0.646199   &      1.064923   &      0.403686   &    H      &      -1.563761   &   2.161296   &   1.612599 \\[-0.3ex]    
   C      &     -1.662128   &      4.335291   &     -0.960511   &    C      &      -1.246382   &  -0.600803   &   0.560203 \\[-0.3ex]    
   C      &      3.988831   &     -2.534923   &     -0.407011   &    H      &      -1.897437   &   0.182781   &   0.938003 \\[-0.3ex]    
   C      &      2.417276   &     -0.273303   &      0.920034   &    H      &      -1.408254   &  -1.473965   &   1.184047 \\[-0.3ex]    
   C      &      2.091988   &      1.028101   &      0.525909   &    C      &      -4.108759   &  -0.652340   &  -0.709728 \\[-0.3ex]    
   C      &     -0.168766   &      2.183683   &     -0.053397   &    H      &      -3.999512   &   0.310996   &  -1.224059 \\[-0.3ex]    
   C      &      2.995735   &      2.159835   &      0.325164   &    H      &      -4.082838   &  -0.456440   &   0.369563 \\[-0.3ex]    
   C      &      0.419141   &     -2.353870   &      3.069580   &    C      &      -6.650259   &  -0.324075   &  -0.707922 \\[-0.3ex]    
   H      &      0.910256   &     -1.637131   &      3.733725   &    H      &      -7.568066   &  -0.730878   &  -1.153274 \\[-0.3ex]    
   H      &     -0.642953   &     -2.101991   &      3.017204   &    H      &      -6.495806   &   0.668359   &  -1.140733 \\[-0.3ex]    
   H      &      0.504902   &     -3.348812   &      3.516319   &    O      &      -2.957802   &  -2.795824   &  -0.381657 \\[-0.3ex]    
   C      &      4.268662   &     -3.909655   &     -0.398697   &    H      &      -3.906909   &  -3.030140   &  -0.297532 \\[-0.3ex]    
   H      &      4.731041   &     -4.357969   &      0.475838   &    O      &      -5.655569   &  -2.545147   &  -0.464068 \\[-0.3ex]    
   C      &      3.948363   &     -4.695585   &     -1.506193   &    H      &      -6.057580   &  -2.315720   &   0.424185 \\[-0.3ex]    
   H      &      4.170540   &     -5.758090   &     -1.488357   &    O      &      -7.306956   &   0.914596   &   1.252421 \\[-0.3ex]    
   C      &      3.335719   &     -4.122419   &     -2.625540   &    N      &       4.375974   &  -1.775586   &   0.733948 \\[-0.3ex]    
   H      &      3.080796   &     -4.735547   &     -3.484122   &    H      &       5.225697   &  -2.077828   &   1.200545 \\[-0.3ex]    
   C      &      3.063944   &     -2.751871   &     -2.633742   &    C      &       1.096393   &  -2.381733   &   1.676075 \\[-0.3ex]    
   H      &      2.604575   &     -2.291613   &     -3.503151   &    H      &       2.130715   &  -2.693067   &   1.845509 \\[-0.3ex]    
   C      &      3.399109   &     -1.952554   &     -1.537433   &    C      &       0.474469   &  -3.448720   &   0.743638 \\[-0.3ex]    
   H      &      3.212727   &     -0.886531   &     -1.576653   &    H      &      -0.597725   &  -3.314107   &   0.589757 \\[-0.3ex]    
   C      &      2.659668   &      3.450942   &      0.792460   &    H      &       0.968987   &  -3.458168   &  -0.231308 \\[-0.3ex]    
   H      &      1.723235   &      3.610824   &      1.314216   &    H      &       0.624404   &  -4.428748   &   1.206316 \\[-0.3ex]    
   C      &      3.544693   &      4.514396   &      0.629841   &    C      &      -1.550181   &  -0.927290   &  -0.908957 \\[-0.3ex]    
   H      &      3.278840   &      5.495825   &      1.009391   &    H      &      -0.797094   &  -1.628479   &  -1.283944 \\[-0.3ex]    
   C      &      4.770850   &      4.317880   &     -0.013977   &    H      &      -1.481698   &  -0.021257   &  -1.519985 \\[-0.3ex]    
   H      &      5.455591   &      5.149702   &     -0.146668   &    C      &      -2.934244   &  -1.558994   &  -1.109676 \\[-0.3ex]    
   C      &      5.115952   &      3.043918   &     -0.484017   &    H      &      -3.027133   &  -1.775205   &  -2.186690 \\[-0.3ex]    
   H      &      6.063778   &      2.887122   &     -0.988938   &    C      &      -5.478501   &  -1.249984   &  -1.079703 \\[-0.3ex]    
   C      &      4.244199   &      1.974154   &     -0.308148   &    H      &      -5.496419   &  -1.422797   &  -2.163334 \\[-0.3ex]    
   H      &      4.520721   &      0.996544   &     -0.687678   &           &                  &              &            \\[-0.3ex]
    \hline 
    \end{tabular*}
  \label{table:ATV1H}
\end{center}
\end{table}
\begin{table} 
\small 
  \begin{center}
    \caption{Coordinates of the \ce{ATVwoH-}  anion optimized in methanol via RB3LYP/6-31+G(d,p)/IEFPCM. All values in angstrom.}
    \begin{tabular*}{0.95\textwidth}{@{\extracolsep{\fill}}rrrr|rrrr}
Atom      &        X      &     Y      &      Z       &   Atom    &       X        &    Y       &      Z     \\
\hline
   F      &    -2.911591  &  5.105072  & -1.340920    &    C      &     -0.232085  &  2.648249  & -1.352553    \\[-0.3ex]    
   O      &    -6.720822  & -1.504172  &  1.466889    &    H      &      0.535068  &  2.273234  & -2.022359    \\[-0.3ex]    
   O      &     4.382747  &  0.194872  &  2.261997    &    C      &     -1.087664  &  3.660917  & -1.791896    \\[-0.3ex]    
   N      &     0.205376  & -0.228705  &  0.736299    &    H      &     -1.003010  &  4.082429  & -2.787733    \\[-0.3ex]    
   C      &    -6.889240  & -0.417880  &  0.809901    &    C      &     -2.219598  &  3.614213  &  0.366233    \\[-0.3ex]    
   C      &     3.812973  & -0.466796  &  1.380006    &    H      &     -2.992024  &  4.006276  &  1.018960    \\[-0.3ex]    
   C      &     1.313476  & -0.982351  &  1.077178    &    C      &     -1.347519  &  2.608868  &  0.790049    \\[-0.3ex]    
   C      &     0.601785  &  1.061922  &  0.395428    &    H      &     -1.438067  &  2.226867  &  1.802274    \\[-0.3ex]    
   C      &    -2.066244  &  4.117581  & -0.918994    &    C      &     -1.167396  & -0.728943  &  0.563444    \\[-0.3ex]    
   C      &     4.227019  & -2.328095  & -0.324417    &    H      &     -1.851053  &  0.032111  &  0.940451    \\[-0.3ex]    
   C      &     2.427252  & -0.153546  &  0.950055    &    H      &     -1.293183  & -1.607299  &  1.190842    \\[-0.3ex]    
   C      &     1.980659  &  1.146037  &  0.535002    &    C      &     -4.049931  & -0.813337  & -0.687171    \\[-0.3ex]    
   C      &    -0.346320  &  2.105699  & -0.060327    &    H      &     -3.934300  &  0.168402  & -1.164223    \\[-0.3ex]    
   C      &     2.819490  &  2.342850  &  0.302488    &    H      &     -4.033999  & -0.656567  &  0.398674    \\[-0.3ex]    
   C      &     0.614082  & -2.434890  &  3.049656    &    C      &     -6.592890  & -0.481722  & -0.706110    \\[-0.3ex]    
   H      &     1.057791  & -1.697797  &  3.726480    &    H      &     -7.502315  & -0.862609  & -1.190284    \\[-0.3ex]    
   H      &    -0.462476  & -2.244038  &  3.003279    &    H      &     -6.426279  &  0.529733  & -1.087672    \\[-0.3ex]    
   H      &     0.753310  & -3.428571  &  3.489389    &    O      &     -2.896403  & -2.960337  & -0.403998    \\[-0.3ex]    
   C      &     4.722031  & -3.642018  & -0.320880    &    H      &     -3.843859  & -3.204448  & -0.346895    \\[-0.3ex]    
   H      &     5.239966  & -4.016380  &  0.557901    &    O      &     -5.603693  & -2.712995  & -0.538912    \\[-0.3ex]    
   C      &     4.548555  & -4.464819  & -1.435147    &    H      &     -6.021070  & -2.520862  &  0.350468    \\[-0.3ex]    
   H      &     4.938823  & -5.478053  & -1.415335    &    O      &     -7.308144  &  0.664780  &  1.288689    \\[-0.3ex]    
   C      &     3.869935  & -3.992784  & -2.562888    &    N      &      4.486062  & -1.527240  &  0.818037    \\[-0.3ex]    
   H      &     3.729360  & -4.633650  & -3.427767    &    H      &      5.352026  & -1.732117  &  1.305552    \\[-0.3ex]    
   C      &     3.381138  & -2.682669  & -2.567024    &    C      &      1.285514  & -2.382753  &  1.658040    \\[-0.3ex]    
   H      &     2.863562  & -2.297507  & -3.440779    &    H      &      2.339857  & -2.623517  &  1.825152    \\[-0.3ex]    
   C      &     3.563300  & -1.846657  & -1.462953    &    C      &      0.743124  & -3.483242  &  0.720946    \\[-0.3ex]    
   H      &     3.191257  & -0.830237  & -1.490106    &    H      &     -0.335493  & -3.411483  &  0.561870    \\[-0.3ex]    
   C      &     2.478766  &  3.587309  &  0.864846    &    H      &      1.236697  & -3.447289  & -0.254633    \\[-0.3ex]    
   H      &     1.593283  &  3.664958  &  1.487858    &    H      &      0.945267  & -4.464415  &  1.164425    \\[-0.3ex]    
   C      &     3.269482  &  4.717889  &  0.645105    &    C      &     -1.489927  & -1.073031  & -0.899286    \\[-0.3ex]    
   H      &     2.985431  &  5.666297  &  1.092733    &    H      &     -0.733571  & -1.770993  & -1.275210    \\[-0.3ex]    
   C      &     4.425984  &  4.629795  & -0.136626    &    H      &     -1.425337  & -0.168901  & -1.514360    \\[-0.3ex]    
   H      &     5.042885  &  5.507704  & -0.305086    &    C      &     -2.870900  & -1.706336  & -1.106308    \\[-0.3ex]    
   C      &     4.781975  &  3.398460  & -0.696254    &    H      &     -2.969820  & -1.903943  & -2.187097    \\[-0.3ex]    
   H      &     5.676763  &  3.315206  & -1.306889    &    C      &     -5.415550  & -1.393226  & -1.097006    \\[-0.3ex]    
   C      &     3.986514  &  2.270621  & -0.480523    &    H      &     -5.418934  & -1.521690  & -2.187140    \\[-0.3ex]    
   H      &     4.271472  &  1.326034  & -0.934231    &           &                &            &              \\[-0.3ex]
    \hline 
    \end{tabular*}
  \label{table:ATV1H-}
\end{center}
\end{table}
\begin{table*} 
\small 
  \begin{center}
    \caption{Coordinates of ascorbic acid and its radical corresponding to the lowest proton affinity optimized in methanol via B3LYP/6-31+G(d,p)/IEFPCM. All values in angstrom.}
    \begin{tabular*}{0.95\textwidth}{@{\extracolsep{\fill}}rrrr|rrr}
                  &                & \ce{C6H8O6} &              & $\ce{C6H7O6^{\bullet -}}$  &   3-OH     &  position    \\
\hline
                  &       X        &   Y         &      Z       &      X        &     Y      &     Z        \\
\hline
            C     &       1.732673 &  -1.121442  &  0.037517    &    1.696486   &  -1.056929 &   0.018792   \\  
            C     &       1.987555 &   0.305517  & -0.092773    &    1.916435   &   0.345380 &   0.065484   \\  
            O     &       0.428463 &  -1.317156  &  0.418174    &    0.347601   &  -1.332120 &   0.195843   \\  
            O     &       2.504586 &  -2.045065  & -0.157602    &    2.509137   &  -1.973278 &  -0.156136   \\  
            C     &       0.850271 &   0.968796  &  0.198893    &    0.720808   &   1.015066 &   0.253760   \\  
            O     &       3.189176 &   0.806773  & -0.477857    &    3.164716   &   0.911835 &  -0.115687   \\  
            C     &      -0.204282 &  -0.034166  &  0.603625    &   -0.342761   &  -0.077749 &   0.393805   \\  
            O     &       0.637153 &   2.291594  &  0.184709    &    0.415105   &   2.253140 &   0.289918   \\  
            C     &      -1.491253 &   0.060295  & -0.223930    &   -1.470505   &   0.092994 &  -0.636161   \\  
            O     &      -3.805847 &  -0.611458  & -0.442660    &   -3.289351   &  -0.757284 &   0.766394   \\  
            C     &      -2.575739 &  -0.922087  &  0.220000    &   -2.610228   &  -0.910782 &  -0.480920   \\  
            O     &      -1.939695 &   1.408401  & -0.028218    &   -2.039638   &   1.396602 &  -0.480203   \\  
            H     &       3.798179 &   0.064143  & -0.623901    &    3.793706   &   0.185007 &  -0.246780   \\  
            H     &      -0.448099 &   0.070101  &  1.668412    &   -0.766450   &  -0.080107 &   1.404618   \\  
            H     &      -0.330233 &   2.444840  &  0.190510    &   -1.296892   &   1.999544 &  -0.232308   \\  
            H     &      -1.249929 &  -0.102852  & -1.283923    &   -1.038960   &  -0.025384 &  -1.642569   \\  
            H     &      -3.799267 &  -1.001417  & -1.328283    &   -3.310756   &  -0.775462 &  -1.316932   \\  
            H     &      -2.262982 &  -1.953322  &  0.027307    &   -2.223541   &  -1.932286 &  -0.514446   \\  
            H     &      -2.768903 &  -0.802305  &  1.290241    &   -3.479077   &   0.189654 &   0.858817   \\  
            H     &      -2.844799 &   1.474624  & -0.372721    &               &            &              \\  
    \hline                                                                                                  
    \end{tabular*}
  \label{table:vitC}
\end{center}
\end{table*}


\begin{thebibliography}{10}
\expandafter\ifx\csname url\endcsname\relax
  \def\url#1{\texttt{#1}}\fi
\expandafter\ifx\csname urlprefix\endcsname\relax\def\urlprefix{URL }\fi
\expandafter\ifx\csname href\endcsname\relax
  \def\href#1#2{#2} \def\path#1{#1}\fi

\bibitem{Roth:02}
B.~D. Roth, The discovery and development of atorvastatin, a potent novel
  hypolipidemic agent, Vol.~40 of Progress in Medicinal Chemistry, Elsevier,
  2002, pp. 1--22.
\newblock \href
  {http://dx.doi.org/https://doi.org/10.1016/S0079-6468(08)70080-8}
  {\path{doi:https://doi.org/10.1016/S0079-6468(08)70080-8}}.

\bibitem{Istvan:01}
E.~S. Istvan, J.~Deisenhofer, Structural mechanism for statin inhibition of
  hmg-coa reductase, Science 292~(5519) (2001) 1160--1164.
\newblock \href {http://dx.doi.org/10.1126/science.1059344}
  {\path{doi:10.1126/science.1059344}}.

\bibitem{Istvan:02}
E.~S. Istvan, Structural mechanism for statin inhibition of
  3-hydroxy-3-methylglutaryl coenzyme a reductase, American Heart Journal
  144~(6, Supplement) (2002) S27--S32.
\newblock \href {http://dx.doi.org/https://doi.org/10.1067/mhj.2002.130300}
  {\path{doi:https://doi.org/10.1067/mhj.2002.130300}}.

\bibitem{Vuppalanchi:06}
R.~Vuppalanchi, N.~Chalasani, Statins for hyperlipidemia in patients with
  chronic liver disease: Are they safe?, Clinical Gastroenterology and
  Hepatology 4~(7) (2006) 838--839.
\newblock \href {http://dx.doi.org/https://doi.org/10.1016/j.cgh.2006.04.020}
  {\path{doi:https://doi.org/10.1016/j.cgh.2006.04.020}}.

\bibitem{Ingold:14}
K.~U. Ingold, D.~A. Pratt, Advances in radical-trapping antioxidant chemistry
  in the 21st century: A kinetics and mechanisms perspective, Chem. Rev.
  114~(18) (2014) 9022--9046, pMID: 25180889.
\newblock \href {http://dx.doi.org/10.1021/cr500226n}
  {\path{doi:10.1021/cr500226n}}.

\bibitem{Alenghat:19}
F.~J. Alenghat, A.~M. Davis, {Management of Blood Cholesterol}, JAMA 321~(8)
  (2019) 800--801.
\newblock \href {http://dx.doi.org/10.1001/jama.2019.0015}
  {\path{doi:10.1001/jama.2019.0015}}.

\bibitem{Galano:19}
A.~Galano, J.~Raul Alvarez-Idaboy, Computational strategies for predicting free
  radical scavengers' protection against oxidative stress: Where are we and
  what might follow?, Int. J. Quantum Chem. 119~(2) (2019) e25665.
\newblock \href {http://dx.doi.org/https://doi.org/10.1002/qua.25665}
  {\path{doi:https://doi.org/10.1002/qua.25665}}.

\bibitem{Duque:22}
L.~Duque, G.~Guerrero, J.~H. Colorado, J.~A. Restrepo, E.~Velez, Theoretical
  insight into mechanism of antioxidant capacity of atorvastatin and its
  o-hydroxy and p-hydroxy metabolites, using dft methods, Comp Theor. Chem.
  (2022) 113758\href
  {http://dx.doi.org/https://doi.org/10.1016/j.comptc.2022.113758}
  {\path{doi:https://doi.org/10.1016/j.comptc.2022.113758}}.

\bibitem{g16}
M.~J. Frisch, G.~W. Trucks, H.~B. Schlegel, G.~E. Scuseria, M.~A. Robb, J.~R.
  Cheeseman, G.~Scalmani, V.~Barone, G.~A. Petersson, H.~Nakatsuji, X.~Li,
  M.~Caricato, A.~V. Marenich, J.~Bloino, B.~G. Janesko, R.~Gomperts,
  B.~Mennucci, H.~P. Hratchian, J.~V. Ortiz, A.~F. Izmaylov, J.~L. Sonnenberg,
  D.~Williams-Young, F.~Ding, F.~Lipparini, F.~Egidi, J.~Goings, B.~Peng,
  A.~Petrone, T.~Henderson, D.~Ranasinghe, V.~G. Zakrzewski, J.~Gao, N.~Rega,
  G.~Zheng, W.~Liang, M.~Hada, M.~Ehara, K.~Toyota, R.~Fukuda, J.~Hasegawa,
  M.~Ishida, T.~Nakajima, Y.~Honda, O.~Kitao, H.~Nakai, T.~Vreven,
  K.~Throssell, J.~J.~A.~Montgomery, J.~E. Peralta, F.~Ogliaro, M.~J. Bearpark,
  J.~J. Heyd, E.~N. Brothers, K.~N. Kudin, V.~N. Staroverov, T.~A. Keith,
  R.~Kobayashi, J.~Normand, K.~Raghavachari, A.~P. Rendell, J.~C. Burant, S.~S.
  Iyengar, J.~Tomasi, Cossi, M.~Millam, Klene, C.~Adamo, R.~Cammi, J.~W.
  Ochterski, R.~L. Martin, K.~Morokuma, O.~Farkas, J.~B. Foresman, D.~J. Fox,
  \href{www.gaussian.com}{Gaussian, inc., wallingford ct, gaussian 16, revision
  b.01} (2016).
\newline\urlprefix\url{www.gaussian.com}

\bibitem{bwHPC2}
bwHPC, bwhpc 2.0 (justus cluster\_2.0, bwunicluster\_2.0, mls\&wiso\_2.0)
  supported by the state of baden-w\"{u}rttemberg supported by the state of
  baden-württemberg and the german research foundation (dfg) through grant no
  inst 40/575-1 fugg. (2020).

\bibitem{Parr:88}
C.~Lee, W.~Yang, R.~G. Parr, Development of the colle-salvetti
  correlation-energy formula into a functional of the electron density, Phys.
  Rev. B 37 (1988) 785--789.
\newblock \href {http://dx.doi.org/10.1103/PhysRevB.37.785}
  {\path{doi:10.1103/PhysRevB.37.785}}.

\bibitem{Becke:88}
A.~D. Becke, Density-functional exchange-energy approximation with correct
  asymptotic behavior, Phys. Rev. A 38 (1988) 3098--3100.
\newblock \href {http://dx.doi.org/10.1103/PhysRevA.38.3098}
  {\path{doi:10.1103/PhysRevA.38.3098}}.

\bibitem{Becke:93a}
A.~D. Becke, A new mixing of hartree-fock and local density-functional
  theories, J. Chem. Phys. 98~(2) (1993) 1372--1377.
\newblock \href {http://dx.doi.org/10.1063/1.464304}
  {\path{doi:10.1063/1.464304}}.

\bibitem{Frisch:94}
P.~J. Stephens, J.~F. Devlin, C.~F. Chabalowski, M.~J. Frisch, Ab initio
  calculation of vibrational absorption and circular dichroism spectra using
  density functional force fields, J. Phys. Chem. 98~(45) (1994) 11623--11627.
\newblock \href {http://dx.doi.org/10.1021/j100096a001}
  {\path{doi:10.1021/j100096a001}}.

\bibitem{Petersson:88}
G.~A. Petersson, A.~Bennett, T.~G. Tensfeldt, M.~A. Al-Laham, W.~A. Shirley,
  J.~Mantzaris, A complete basis set model chemistry. i. the total energies of
  closed-shell atoms and hydrides of the first-row elements, J. Chem. Phys.
  89~(4) (1988) 2193--2218.
\newblock \href {http://dx.doi.org/10.1063/1.455064}
  {\path{doi:10.1063/1.455064}}.

\bibitem{Petersson:91}
G.~A. Petersson, M.~A. Al-Laham, A complete basis set model chemistry. ii.
  open-shell systems and the total energies of the first-row atoms, J. Chem.
  Phys. 94~(9) (1991) 6081--6090.
\newblock \href {http://dx.doi.org/10.1063/1.460447}
  {\path{doi:10.1063/1.460447}}.

\bibitem{Tomasi:05}
J.~Tomasi, B.~Mennucci, R.~Cammi, Quantum mechanical continuum solvation
  models, Chem. Rev. 105~(8) (2005) 2999--3094, pMID: 16092826.
\newblock \href {http://dx.doi.org/10.1021/cr9904009}
  {\path{doi:10.1021/cr9904009}}.

\bibitem{Cances:97}
E.~Canc\`es, B.~Mennucci, J.~Tomasi, A new integral equation formalism for the
  polarizable continuum model: Theoretical background and applications to
  isotropic and anisotropic dielectrics, J. Chem. Phys. 107~(8) (1997)
  3032--3041.
\newblock \href {http://dx.doi.org/10.1063/1.474659}
  {\path{doi:10.1063/1.474659}}.

\bibitem{Baldea:2019e}
I.~B\^aldea, Long carbon-based chains of interstellar medium can have a triplet
  ground state. why is this important for astrochemistry?, ACS Earth Space
  Chem. 3~(5) (2019) 863--872.
\newblock \href {http://dx.doi.org/10.1021/acsearthspacechem.9b00008}
  {\path{doi:10.1021/acsearthspacechem.9b00008}}.

\bibitem{Baldea:2019g}
I.~B\^aldea, Alternation of singlet and triplet states in carbon-based chain
  molecules and its astrochemical implications: Results of an extensive
  theoretical study, Adv. Theor. Simul. 2~(9) (2019) 1900084.
\newblock \href {http://dx.doi.org/10.1002/adts.201900084}
  {\path{doi:10.1002/adts.201900084}}.

\bibitem{Baldea:2022d}
I.~B\^aldea, \ce{HCnH-} anion chains with $n\leq 8$ are nonlinear and their
  permanent dipole makes them potential candidates for astronomical
  observation, Molecules 27~(10).
\newblock \href {http://dx.doi.org/https://doi.org/10.3390/molecules27103100}
  {\path{doi:https://doi.org/10.3390/molecules27103100}}.

\bibitem{Baldea:2022e}
I.~B\^aldea, Comprehensive quantum chemical characterization of the
  astrochemically relevant \ce{HC_nH} chain family. an attempt to aid
  astronomical observations, Adv. Theor. Simul. DOI:10.1002/adts.202200244
  (2022) DOI:10.1002/adts.202200244.
\newblock \href {http://dx.doi.org/DOI:10.1002/adts.202200244}
  {\path{doi:DOI:10.1002/adts.202200244}}.

\bibitem{gabedit}
A.-R. Allouche, Gabedit: A graphical user interface for computational chemistry
  softwares, J. Comput. Chem. 32~(1) (2011) 174--182.
\newblock \href {http://dx.doi.org/10.1002/jcc.21600}
  {\path{doi:10.1002/jcc.21600}}.

\bibitem{Gunnarson:89}
R.~O. Jones, O.~Gunnarsson, The density functional formalism, its applications
  and prospects, Rev. Mod. Phys. 61 (1989) 689--746.
\newblock \href {http://dx.doi.org/10.1103/RevModPhys.61.689}
  {\path{doi:10.1103/RevModPhys.61.689}}.

\bibitem{Baldea:2012i}
I.~B\^aldea, Extending the newns-anderson model to allow nanotransport studies
  through molecules with floppy degrees of freedom, Europhys. Lett. 99~(4)
  (2012) 47002.
\newblock \href {http://dx.doi.org/10.1209/0295-5075/99/47002}
  {\path{doi:10.1209/0295-5075/99/47002}}.

\bibitem{Baldea:2014c}
I.~B\^aldea, A quantum chemical study from a molecular transport perspective:
  Ionization and electron attachment energies for species often used to
  fabricate single-molecule junctions, Faraday Discuss. 174 (2014) 37--56.
\newblock \href {http://dx.doi.org/10.1039/C4FD00101J}
  {\path{doi:10.1039/C4FD00101J}}.

\bibitem{Fifen:13}
J.~J. Fifen, Thermodynamics of the electron revisited and generalized, Journal
  of Chemical Theory and Computation 9~(7) (2013) 3165--3169, pMID: 26583993.
\newblock \href {http://dx.doi.org/10.1021/ct400212t}
  {\path{doi:10.1021/ct400212t}}.

\bibitem{Fifen:14}
J.~J. Fifen, Z.~Dhaouadi, M.~Nsangou, Revision of the thermodynamics of the
  proton in gas phase, The Journal of Physical Chemistry A 118~(46) (2014)
  11090--11097, pMID: 25338234.
\newblock \href {http://dx.doi.org/10.1021/jp508968z}
  {\path{doi:10.1021/jp508968z}}.

\bibitem{Markovic:16}
Z.~Markovic, J.~Tosovic, D.~Milenkovic, S.~Markovic, Revisiting the solvation
  enthalpies and free energies of the proton and electron in various solvents,
  Computational and Theoretical Chemistry 1077 (2016) 11--17, antioxidants vs.
  Oxidative Stress: Insights from Computation.
\newblock \href
  {http://dx.doi.org/https://doi.org/10.1016/j.comptc.2015.09.007}
  {\path{doi:https://doi.org/10.1016/j.comptc.2015.09.007}}.

\bibitem{Rimarcik:10}
J.~Rimarcik, V.~Lukes, E.~Klein, M.~Ilcin, Study of the solvent effect on the
  enthalpies of homolytic and heterolytic n-h bond cleavage in
  p-phenylenediamine and tetracyano-p-phenylenediamine, Journal of Molecular
  Structure: THEOCHEM 952~(1) (2010) 25--30.
\newblock \href
  {http://dx.doi.org/https://doi.org/10.1016/j.theochem.2010.04.002}
  {\path{doi:https://doi.org/10.1016/j.theochem.2010.04.002}}.

\bibitem{Parr:89}
R.~G. Parr, W.~Yang, Density-Functional Theory of Atoms and Molecules, Oxford
  University Press, Clarendon, Oxford, 1989, see p.~149.

\bibitem{Burke:12}
K.~Burke, Perspective on density functional theory, J. Chem. Phys. 136~(15)
  (2012) 150901.
\newblock \href {http://dx.doi.org/http://dx.doi.org/10.1063/1.4704546}
  {\path{doi:http://dx.doi.org/10.1063/1.4704546}}.

\bibitem{Perdew:82}
J.~P. Perdew, R.~G. Parr, M.~Levy, J.~L. Balduz, Density-functional theory for
  fractional particle number: Derivative discontinuities of the energy, Phys.
  Rev. Lett. 49 (1982) 1691--1694.
\newblock \href {http://dx.doi.org/10.1103/PhysRevLett.49.1691}
  {\path{doi:10.1103/PhysRevLett.49.1691}}.

\bibitem{Baldea:2013c}
I.~B\^aldea, Demonstrating why dft-calculations for molecular transport in
  solvents need scissor corrections, Electrochem. Commun. 36 (2013) 19--21.
\newblock \href
  {http://dx.doi.org/http://dx.doi.org/10.1016/j.elecom.2013.08.027}
  {\path{doi:http://dx.doi.org/10.1016/j.elecom.2013.08.027}}.

\end{thebibliography}
\end{document}